\documentclass{article}

\usepackage{arxiv}

\usepackage[utf8]{inputenc} 
\usepackage[nottoc]{tocbibind}
\usepackage[T1]{fontenc}    
\usepackage{hyperref}       
\usepackage{url}            
\usepackage{booktabs}       
\usepackage{amsfonts}       
\usepackage{nicefrac}       
\usepackage{microtype}      
\usepackage{cite}
\usepackage[table]{xcolor}
\usepackage{amsmath,amssymb,amsfonts}
\usepackage{algorithmic}
\usepackage{graphicx}
\usepackage{textcomp}
\usepackage{xcolor}
\usepackage{hyperref}
\usepackage{nicefrac}
\usepackage[T1]{fontenc}
\usepackage{microtype}
\usepackage[english]{babel}
\usepackage{newtxmath}
\usepackage[scaled]{helvet} 
\usepackage{cite}
\usepackage{amsmath,amssymb,amsfonts}
\usepackage{nicefrac}
\usepackage{paralist}
\usepackage{algorithmic}
\usepackage{graphicx}
\usepackage{textcomp}
\usepackage{stfloats}
\usepackage{xcolor}
\usepackage{balance}
\usepackage{paralist}
\usepackage{algorithmic}
\usepackage{graphicx}
\usepackage{textcomp}
\usepackage[table]{xcolor}
\usepackage{balance}
\usepackage{enumitem}
\usepackage{booktabs}
\usepackage{multirow}
\usepackage{hhline}
\usepackage{url}
\usepackage{tcolorbox}
\usepackage{xspace}
\usepackage{stfloats}
\def\BibTeX{{\rm B\kern-.05em{\sc i\kern-.025em b}\kern-.08em
    T\kern-.1667em\lower.7ex\hbox{E}\kern-.125emX}}
\usepackage{paralist}

\title{Establishing a Search String to Detect Secondary Studies in Software Engineering}

\author{
Bianca M. Napole\~ao \\
Université du Québec à Chicoutimi \\
Chicoutimi, QC, Canada \\
\texttt{bianca-minetto.napoleao1@uqca.ca} 
\And
Katia R. Felizardo, \'Erica F. de Souza \\
Federal University of Technology -- Paran\'a \\
Cornélio Procópio, PR, Brazil \\
\texttt{katiascannavino, ericasouza@utfpr.edu.br}
\And
Fabio Petrillo, Sylvain Hall\'e\\
Université du Québec à Chicoutimi \\
Chicoutimi, QC, Canada \\
\texttt{fabio@petrillo.com, shalle@acm.org} 
\And
Nandamudi L. Vijaykumar\\
National Inst. for Space Research \\
São José dos Campos, SP, Brazil \\
\texttt{vijay.nl@inpe.br} 
\And
Elisa Y. Nakagawa\\
University of S\~ao Paulo\\
S\~ao Paulo, SP, Brazil \\
\texttt{elisa@icmc.usp.br} 
}

\date{}
\begin{document}
\maketitle

\begin{abstract}
 \textbf{Context:} A tertiary study can be performed to identify related reviews on a topic of interest. 
    However, the elaboration of an appropriate and effective search string to detect secondary studies is challenging for Software Engineering (SE) researchers.
    \textbf{Objective:} The main goal of this study is to propose a suitable search string to detect secondary studies in SE, addressing issues such as the quantity of applied terms, relevance, recall and precision. \textbf{Method:} We analyzed seven tertiary studies under two perspectives: (1) structure -- strings' terms to detect secondary studies; and (2) field: where searching -- titles alone or abstracts alone or titles and abstracts together, among others. We validate our string by performing a two-step validation process. Firstly, we evaluated the capability to retrieve secondary studies over a set of 1537 secondary studies included in 24 tertiary studies in SE. Secondly, we evaluated the general capacity of retrieving secondary studies over an automated search using the Scopus digital library. \textbf{Results:} Our string was capable to retrieve an optimum value of over 90\% of the included secondary studies (recall) with a high general precision of almost 60\%.  \textbf{Conclusion:} The suitable search string for finding secondary studies in SE contains the terms ``systematic review'', ``literature review'', ``systematic mapping'', ``mapping study'' and ``systematic map''.
\end{abstract}


\section{Introduction}\label{sec:introduction}

Kitchenham \textit{et al.} \cite{Kitchenham15} state that a tertiary study can be performed to identify related reviews on the topic of interest for categorizing and observing research trends.
A tertiary study is a kind of SLR in which the inputs are secondary studies: Systematic Literature Reviews (SLRs) and Systematic Mappings (SMs). There are many challenges associated with how it is conducted, including the search for the inputs. In addition, it is unclear how to decide which terms to include in the search string to find secondary studies \cite{Zhang11a}. 



Software Engineering (SE) researchers have been adopting different terms for searching secondary studies, e.g., the search string used in Silva \textit{et al.} \cite{Silva10} tertiary study contains 18 secondary studies' related terms while in Hanssen \textit{et al.} \cite{Hanssen11} only 5 secondary studies' terms are present. There is no consensus either on the preferable field(s) to search for secondary studies, e.g. in Cruzes \textit{et al.} \cite{Cruzes11}, title alone was used, but in Delavari \textit{et al.} \cite{Delavari2019}, title, abstract and keywords were used together.

It is difficult to determine when the search string is complete to perform the search. The use of a broader search string could help reviewers during the search for studies. 
Indeed, a broader search string with few terms is easier to adapt. However, choosing few terms can result in loss of evidence because some important terms could be missing. Moreover, a search string can be formed by different combinations. 
Therefore, it is essential to verify which terms have real impact in detecting relevant secondary studies and also eliminate terms that reduce search accuracy and do not aggregate relevant studies to the performing study. 

The definition of a search string is a time-consuming and error-prone activity. A hindrance is the lack of formalization of the terminologies in most of the research topics/domains in which SLRs have been conducted. 
In general, the probability that two researchers use the same term to refer to the same concept is often lower than 20\% and there is no common terminology and appropriate descriptors and keywords in the SE area \cite{felizardo16}. 

In this study, we propose a suitable search string to detect secondary studies in SE, addressing issues such as the number of applied terms, relevance, \textit{recall} and \textit{precision}.
We validated the results through a two-step validation process. Firstly, we evaluated the capability to retrieve secondary studies over a set of 1537 secondary studies included in 24 tertiary studies in SE. Secondly, we evaluated the general capacity of retrieve SE secondary studies over an automated search using \textit{Scopus} Digital Library (DL). 
Our string was capable to retrieve over 90\%  of the included secondary studies (\textit{recall}) with a general \textit{precision} of almost 60\%. These results compared to search strategies scales used for evaluating search terms \cite{Dieste07} indicates our \textit{recall} rate as optimum (80-99\%) and \textit{precision} as high (25-60\%).

As main related work, Dieste \textit{et al.} \cite{Dieste09} also evaluated \textit{recall} and \textit{precision} of search strategies to find an optimum strategy. Their analysis was performed considering the primary study term ``experiment'' and its synonyms while our analysis considered terms related to secondary studies. They concluded that a search strategy can not return 100\% of the relevant studies, due to a lack of standardization in SE terminology. In \cite{Dieste07}, the authors state that a optimal search strategy strikes a balance between high \textit{recall} and  high \textit{precision}. Our results corroborate both points.

The remainder of this study is organized as follows: Section \ref{methodology} details the study design applied to evaluate search strings for detecting secondary studies. Sections \ref{RQ1} and \ref{RQ2} answer the research questions. Section \ref{validationsec} presents the validation process and its results. Section \ref{discussions} discusses our results. 
Finally, Section \ref{conclusions} concludes our work.

\section {Study Design}
\label{methodology}

In this Section, we present the key aspects of the study design.

\subsection{Selecting the Dataset}
\label{dataset}
In order to achieve our study goal, we based on the study performed by Garousi and M\"antyl\"a \cite{Garousi16}, that presents a list of tertiary studies in SE. We chose this study because it was the first study compiled in a single list of 10 tertiary studies in SE. 
Details and references about the list of the identified tertiary studies from Garousi and M\"antyl\"a's study and the adopted search string of each tertiary study is partially described in a Supplementary Table  document (Tables I and II) available online\footnotemark \footnotetext{\url{https://bit.ly/3d4u74Z}}, i.e. the terms used to detect relevant secondary studies in SE.  We elected to use these lists as the initial basis for conducting our study.

Out of the 10 tertiary studies identified by Garousi and M\"antyl\"a \cite{Garousi16}, we focused on seven studies (ID S2 \cite{Kitchenham10}, S3 \cite{Silva10}, S6 \cite{Cruzes11}, S7 \cite{Hanssen11}, S8 \cite{Marques12}, S9 \cite{Imtiaz13} and S10 \cite{Verner14}). These studies were selected for two main reasons: (i)~they were conducted and double-checked by reviewers with experience in conducting secondary studies; and (ii)~they contained all necessary data to be used in our analyses: the list of included studies, description of search strategy and search string (in case of automated search adoption). Considering that our goal is to analyze string terms, the use of automated search in the study is essential. Consequently, we excluded study S1 because it just used manual search. Study S4 used in its research data collected in another tertiary study \cite{Kitchenham10}, which was already included in our analysis. S5 was excluded because the included secondary studies in its list are not available.

\subsection{Research Questions}
\label{RQ}

To facilitate understanding, we translated our research goal into two Research Questions (RQs):
\begin{itemize}
    \item \textbf{RQ1:} \textit{Which terms should be used to detect secondary studies in SE?} 
    \item \textbf{RQ2:} \textit{What is(are) the preferable field(s) to search for secondary studies: (e.g.\ titles alone, abstracts alone, or titles and abstracts together, among others)?}
\end{itemize}

We converted our RQs into two perspectives in order to analyze data and answer the RQs. The perspectives are detailed next. 

\noindent\textbf{-- Perspective 1 - Structured analysis (RQ1):} The focus was to identify terms and synonyms that compose search strings created to detect secondary studies in SE. 
The structured analysis was performed in two steps. During Step 1.1, we considered string terms related to secondary studies. Terms related to research domain (e.g. ``software testing'', ``agile'', etc.)  were not considered since they are not relevant for our analysis and they do not impact secondary studies terms results. In Step 1.2, these terms were organized in a list to reveal the terms and their synonyms most used by researchers to detect secondary studies. The results are presented in Section \ref{RQ1}.

\noindent\textbf{-- Perspective 2 - Search field (RQ2):} The focus was to identify preferable field(s) (search in title alone, or abstract alone, or title and abstract together, etc.) to detect secondary studies in SE. The analysis was executed in two steps. Initially, during Step 2.1, we listed the secondary studies included in each tertiary study and downloaded all of them. A total of 337 were detected, however one study was not available for download, totaling 336 studies. Then, in Step 2.2, we checked the occurrence of string terms in the title, abstract and keywords of each secondary study. The results are showed in Section \ref{RQ2}. 

We used a data extraction form to extract data from the selected studies and be able to answer our RQs. The extracted data is available online\footnotemark[\value{footnote}]. 

\section{RQ1: Which terms should be used to detect secondary studies in SE?} \label{RQ1}

\begin{figure}[!h]
	\centering
	\includegraphics[width= 0.8\linewidth]{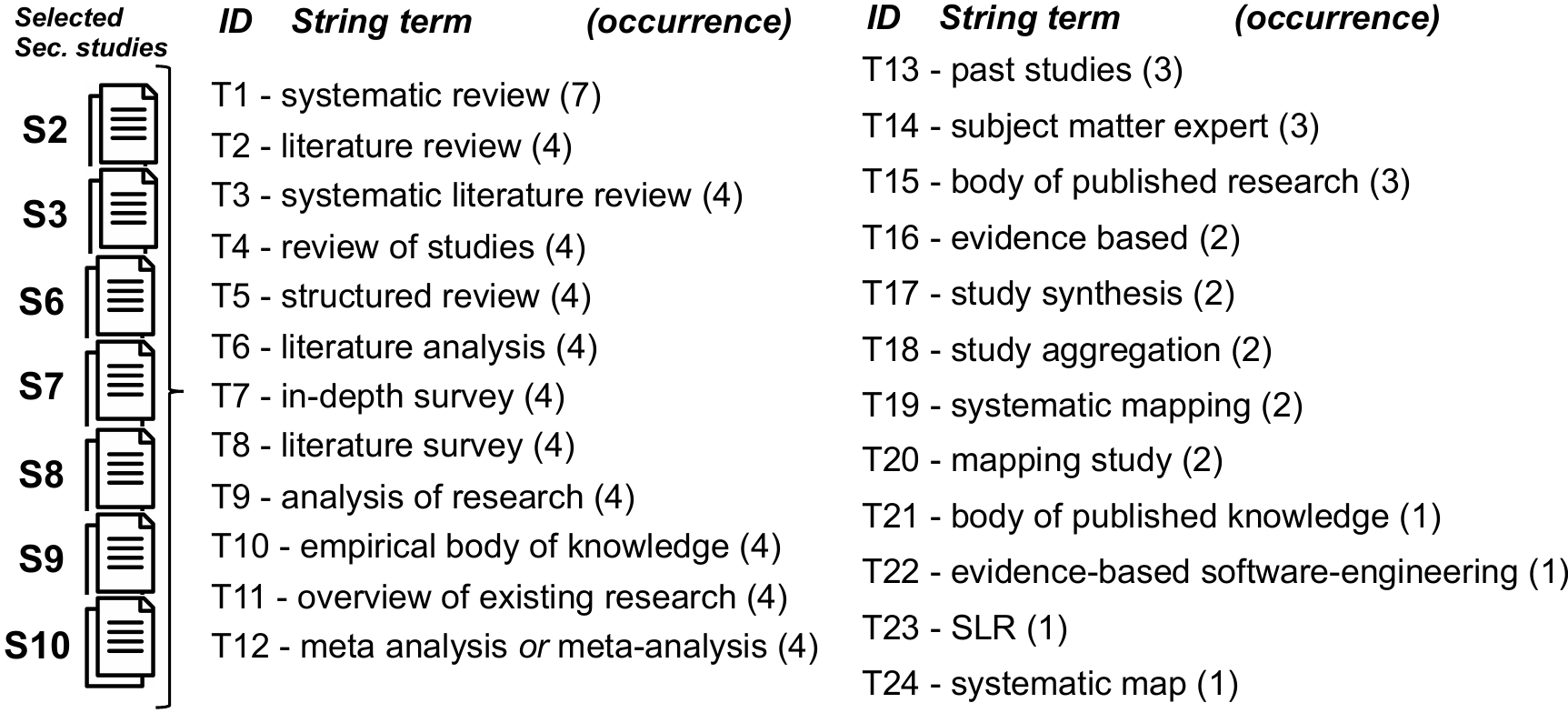}
	\caption{Terms from tertiary studies string sorted by occurrences number.}
	\label{terms}
\end{figure}

Considering the seven tertiary studies selected by us, we extracted and grouped in a single list all search string terms used to find secondary studies. As mentioned in Section \ref{methodology} (Perspective 1), we just considered terms related to secondary studies. These terms were counted and their number of occurrences are listed in Figure \ref{terms}. 

The terms ``meta analysis'' and  ``meta-analysis'' were considered as a unique term (T12), 
since according to Singh and Singh \cite{Singh17}  the character ``-'' is interpreted as a blank space by several DLs, such as \textit{IEEE Xplore}, \textit{ACM DL}, \textit{SpringerLink}, \textit{Science Direct} and \textit{Wiley}.

The most used term in tertiary studies was ``systematic review'' (T1). This term appeared in all seven tertiary strings (100\%). The terms ``literature review'' (T2) and ``systematic literature review'' (T3) were used in four search strings (57.1\%) individually. However, the term ``literature review'' is part of the term ``systematic literature review'' and in six (85.7\%) of the seven tertiary studies, at least one of these terms composed the string. Moreover, in the search strings of studies S9 and S10, both terms (T2 and T3) appeared together. Terms related to systematic mapping (T19, T20, T24) had a lower number of occurrences than terms related to SLR (T1, T2, T3). Similarly, there were less synonyms for SM than for SLR. 


\begin{table*}
	\centering
	\scriptsize
	\caption{String terms that returned secondary studies.}
	\label{tbl:SEterms}
	\begin{tabular}{|p{0.6cm}|p{3.5cm}|c|c|c|c|c|c|c|c|}
		\hline
		\multicolumn{2}{|c|}{\textbf{{ID}}}   & \textbf{{S2}} & \textbf{{S3}} & \textbf{{S6}} & \textbf{{S7}} & \textbf{{S8}} & \textbf{{S9}} & \textbf{{S10}} & \textbf{{Totals}} \\ \hline
		\multicolumn{2}{|c|}{\textbf{\begin{tabular}[c]{@{}c@{}}{Secondary Studies} \end{tabular}}} & {33}          & {67}          & {49}          & {12}          & {23}          & {116}         & {36}           & \textbf{{336}}     \\ \hline
		\multirow{2}{*}{\textbf{{Terms}}}                                & \textbf{{T1: systematic review}}  & {\textbf{9(27.3\%)}}  & \textbf{{41(61.2\%)} }& \textbf{{37(75.5\%)}} & \textbf{{0(0.0\%)} }  & \textbf{{7(30.4\%)} } & \textbf{{67(57.8\%)}} &\textbf{{13(36.1\%)}} & {\textbf{174}}    \\ \cline{2-10} 
		& \textbf{{T2: literature review}}                                & \textbf{8(24.2\%) } & \textbf{{26(38.8\%)} }& {0(0\%)} & {\textbf{4(33.3\%)}}  & {0(0\%)}  & \textbf{{50(43.1\%)} }& \textbf{{15(41.6\%)} } & {\textbf{103}}     \\ \cline{2-10} 
		& \textbf{{T3: systematic literature review}}                                & {0(0\%)}   & {0(0\%)}   & {0(0\%)}   & \textbf{{4(33.3\%)}}  & {\textbf{10(43.5\%)}} & \textbf{{46(39.7\%)}} &\textbf{ {13(36.1\%)}}  & {\textbf{73}}   \\ \cline{2-10} 
		& \textbf{{T4: review of studies}}                               & \textbf{{0(0\%)}}   &\textbf{ {2(3\%)} }  & {0(0\%)}   & {0(0\%)}   & {0(0\%)}   & \textbf{{3(2.6\%)}}   & \textbf{{0(0\%)} }   & {\textbf{5}}       \\ \cline{2-10} 
		& \textbf{{T5: structured review}}                                & \textbf{0(0\%) }  & \textbf{{1(1.5\%)} }  & {0(0\%)}   & {0(0\%)}   & {0(0\%)}  & \textbf{{0(0\%)} }  & \textbf{{0(0\%)} }   & \textbf{{1}}       \\ \cline{2-10} 
		& \textbf{{T6: literature analysis}}                                & \textbf{1(3.0\%)}   & \textbf{{0(0\%)}}  & {0(0\%)}   & {0(0\%)}   & {0(0\%)}   & \textbf{{0(0\%)}}   & \textbf{{1(2.8\%)} }   & \textbf{{2}}       \\ \cline{2-10} 
		& \textbf{{T8: literature survey}}                                & \textbf{1(3.0\%)}   & \textbf{{2(3\%)}}   & {0(0\%)}   & {0(0\%)}   & {0(0\%)}   & \textbf{{2(1.7\%)}}   & \textbf{{0(0\%)} }   & \textbf{{5}}       \\ \cline{2-10} 
		& \textbf{{T9: analysis of research}}                                & \textbf{1(3.0\%)}   & \textbf{{1(1.5\%)}}   & {0(0\%)}   & {0(0\%)}   & {0(0\%)}   & \textbf{{0(0\%)} }  & \textbf{{1(2.8\%)} }   & \textbf{{3}}       \\ \cline{2-10} 
		& \textbf{{T12: meta/meta-analysis}}                               & \textbf{{2(6.0\%)}}   & \textbf{{1(1.5\%)}}   & {0(0\%)}   & {0(0\%)}   & {0(0\%)}   & \textbf{{3(2.6\%)}}   & \textbf{{0(0\%)}}  & \textbf{{6}}       \\ \cline{2-10} 
		& \textbf{{T16: evidence based}}                               & \textbf{0(0\%)}   & \textbf{{4(5.9\%)} }  & {0(0\%)}   & {0(0\%)}   & {0(0\%)}   & \textbf{{24(20.7\%)}} & {0(0.\%)}    & \textbf{{28}}      \\ \cline{2-10} 
		& \textbf{{T23: SLR}}                               & {0(0\%)}   & {0(0\%)}   & {0(0\%)}   & {0(0\%)}   & {0(0\%)}   & \textbf{{19(16.4\%)}} & {0(0\%)}    & \textbf{{19}}      \\ \cline{2-10} 
		& \textbf{{T24: systematic map}}                              & {0(0\%)}   & {0(0\%)}   & {0(0\%)}   & \textbf{{0(0\%)}}   & \textbf{{1(4.3\%)}}   & {0(0\%)}   & {0(0\%)}    & \textbf{{1}}       \\ \hline
	\end{tabular}
\end{table*}

From the analysis presented in Table \ref{tbl:SEterms}, it is possible to identify string terms that returned secondary studies. Terms T7, T10, T11, T13, T14, T15, T17, T18, T19, T20, T21 and T22 are synonyms to SLR; however, in the SE context, they are not being used for investigations to describe their reviews. For this reason, they are not described in Table \ref{tbl:SEterms}. In addition, the bolded values of the study's terms refer to terms present in the referent study's string. 
For example, the tertiary study S2 analyzed 33 secondary studies and the string adopted by S2 was composed by terms that returned studies T1, T2, T6, T8, T9, T12, but also by the terms T4, T5, T7, T10, T11, T15 and T16 which did not return studies. Term T1 (see Table \ref{tbl:SEterms} -- line 3) returned nine of the 33 studies (27.3\%). On one hand, T3 did not find studies because it was not part of the string (see Table \ref{tbl:SEterms} -- line 3). On the other hand, although T4 composed the string, it did not return studies. 


T1 returned 174 (51.8\%) studies of the 336 secondary studies detected by the seven tertiary studies altogether. A total of 103 (30.6\%) studies are returned by term T2 and 73 (21.7\%) was returned using term T3.

One point to be observed is that the same secondary study can be returned by more than one term in the fields. Therefore, the sum of included secondary studies may exceed 100\%. For example, the total number of included studies in S3 is 67, however the sum returned by all terms is 77 (T1 = 41 + T2 = 26 + T4 = 2 + T5 = 1 + T8 = 2 + T9 = 1 + T6 = 4). Similarly, studies returned by combining T1, T2 and T3 amounted to 350 studies ($174 + 103 + 73$). 

Another point to be considered is that the selected tertiary studies S2, S6, S7 and S8 presented in Table \ref{tbl:SEterms} did not solely use an automated search strategy. They also combined complementary search strategies such as manual search \cite{Kitchenham15} and \textit{Snowballing} search \cite{Wohlin14}. Due to this fact, the sum of the included secondary studies by each tertiary study amounts to less than the total of the included secondary studies actually provided by the tertiary study. For example, S6 has 49 included secondary studies, but only 37 of them could be found by the string term. We deduce from this that the remaining 12 studies were added manually.



Considering the results described in this Section, we summarized the synonyms with the top-three higher \textit{recall} of each of the two existing types of secondary studies (SLR and SM) and condensed them in a search string with these terms connected by the logical operator \textit{OR}. Since the term ``literature review'' is capable to return the same studies returned by the term ``systematic literature review'', we considered in our search string only the term ``literature review''.
Accordingly, the search string to detect secondary studies in SE should contain the following secondary studies' terms: \textbf{ ``systematic review'' \textit{OR} ``literature review'' \textit{OR} ``systematic mapping'' \textit{OR} ``mapping study'' \textit{OR} ``systematic map''}. Therefore, a suitable search string to detect secondary studies in SE must consists of the domain terms connected by the logical operator \textit{\textit{AND}} to secondary studies' terms. 

\section{RQ2: What is(are) the preferable field(s) to search secondary studies?} \label{RQ2}


Table \ref{tbl:presence} describes where (title, abstract and keywords) each string term was found in the list of included secondary studies. In order to obtain a global overview from all secondary studies included in the list of all tertiary studies considered, we searched for duplicate studies among the lists of secondary studies as well as analyzed if one respective term was already analyzed before in duplicate papers. For example, studies S2 and S4 had, in their included secondary studies list, the same secondary study and both had the term ``systematic review'' several times appearing in the abstract, the study is counted once as well as the term appearance in the abstract field. This approach was adopted to avoid bias (e.g. wrong relevance of a search field) of  in our overall analysis.

\begin{table} [!h]
	\centering
	\footnotesize
	\caption{Terms appearance on title, abstract and keywords.}
	\label{tbl:presence}
	\begin{tabular}{|l|c|c|c|}
		\hline
		\textbf{{String Terms}} & \textbf{{Title}} & \textbf{{Abstract}} & \textbf{{Keywords}} \\ \hline
		{literature review}                           & {37 }            & {67}                & {43}                \\ \hline
		{systematic review}                           & {82 }           & {95}                & {57}                \\ \hline
		{systematic literature review}                & {31}             & {42}                & {44}                \\ \hline
		{evidence based}        						& 4              & {16}                & {15}                \\ \hline
		{SLR}                                         & {17}             & {1}                 & {2}                 \\ \hline
		{review of studies}                           & {1}              & {3}                 & {0}                 \\ \hline
		{literature survey}                           & {1}              & {4}                 & {1}                 \\ \hline
		{meta[-]analysis}    & {1}             & {4}                 & {2}                 \\ \hline
		{literature analysis}                         & {0}              & {1}                 & {1}                \\ \hline
		{analysis of research}                        & {2}              & {1}                 & {0}                 \\ \hline
		{structured review}                           & {1}              & {0}                 & {1}                 \\ \hline
		{systematic map}                              & {1}              & {0}                 & {1}                 \\ \hline
		{\textbf{TOTAL}}                              & {{\textbf{178}}}              & {{\textbf{234}}}                & {\textbf{167}}                \\ \hline
	\end{tabular}
\end{table}

With respect to the 12 terms that returned secondary studies (Table  \ref{tbl:presence}), abstracts have a higher occurrence of terms, totaling 234 occurrences; this is followed by titles with 178 and keywords with 167. One possible reason is that abstracts present a summary of the study containing more words than title and keywords, consequently the probability of the occurrence of relevant terms is higher than in titles and keywords.
Nevertheless, the presence of some terms in the title were interesting. The term ``SLR'' appeared more often in the title field, while the term ``systematic review'' occurred more often in abstracts (see Table \ref{tbl:presence} -- lines 6 and 3, respectively). We believe this fact occurs because title tends to be more concise than abstract. The term ``systematic literature review'' occurred more often in the list of keywords (see Table \ref{tbl:presence} -- line 4). 


Despite SE authors differ the search fields considered in their searches, we are aware that the results obtained was somehow already expected. However, our analysis confirm that the preferable fields to detect secondary studies in SE are \textbf{title}, \textbf{abstract} and \textbf{keywords} all together. 



\section{Search String and Search Fields Validation} 
\label{validationsec}
This section validates the results obtained for RQ1 and RQ2. 
In order to validate our results, we proposed a two-step validation process. The first step enable the calculation of the \textit{recall} of our string based on the capacity of our string to detect secondary studies from the list of secondary studies included by tertiary studies. The second step enable the calculation of the \textit{precision} of our string based on a the execution of the proposed search string in a DL. We opted to perform the validation process in two independent steps since we do not have access to the list of returned studies by the DLs from each tertiary study considered in our first-step analysis. In the context of our study, \textit{recall} is a metric of ``completeness'' then  \textit{recall} =\begin{math} \frac{returned \ secondary \ studies \ by \ the \ proposed \ string}{included \ secondary \ studies \ by \ tertiary \ studies} \end{math} (see Sections \ref{valprocess1} and \ref{validationresult1}) and \textit{precision} is a metric of ``effort'' then \textit{precision} = \begin{math} \frac{selected \ secondary\ studies \ in \ SE}{retrieved \ studies \ by \ the \ DL} \end{math} (see Sections \ref{valprocess2} and \ref{validationresult2}).

\subsection{Validation process definition -- Step 1}
\label{valprocess1}

The search string suggested by us (RQ1) was tested for retrieving secondary studies. For validation purposes, we defined a process which includes a definition of a control group. The creation of a control group was essential for the evaluation and calibration of the string. When relevant publications of the control group were not found, new terms can be added to the search string during the calibration activity. In order to define the control group, we performed an automated search on the most renowned SE DLs \cite{Zhang11}: 
\textit{Scopus}, \textit{Web of Science}, \textit{IEEE Xplore} and the \textit{ACM Digital Library}. We used the search string: ((\textit{``tertiary study''} OR \textit{``tertiary review''} OR \textit{``tertiary systematic review''} OR  \textit{``systematic review of systematic review''}) AND \textit{``software engineering''}) applied on title, abstract and keywords. The search string was based on the reading of the identified tertiary studies by Garousi and M\"antyl\"a \cite{Garousi16}.
As a result of the automated search, 786 potential tertiary studies were identified. After removing duplicated studies, 713 studies were left.

The next step was to verify if each study met the following selection criteria: (i)~the study is a tertiary study (it must follow renowned guidelines \cite{Kitchenham15, Petersen15} 
and consider secondary studies in its analysis); (ii)~the study should be within the SE context; and (iii)~the list of secondary studies included in the tertiary study is available (essential data to execute our validation process). We excluded a study if it was just published as an abstract, was not written in English, was an older version of some other study already considered, was not in the scope of SE, or when the list of secondary studies it included was unavailable. Our final list of included tertiary studies for the validation process contains 24 tertiary studies published between the years 2012 and 2019. Table III -- Supplementary Tables document\footnotemark[\value{footnote}] presents the list of included tertiary studies for the validation process.

We then constructed a control group of 1537 secondary studies based on the list of included secondary studies of the 24 tertiary studies selected after the application of the selection criteria previously mentioned (see Table 2 -- Supplementary tables document\footnotemark[\value{footnote}]). The sum of the included secondary studies (column 5 -- Table III) presented in the Supplementary Tables document\footnotemark[\value{footnote}] represents our control group.


In order to determine if the proposed search string was able to detect the 1537 secondary studies from our control group, we checked the presence of the terms defined in our string in the title of the studies. Only if the term is  not found in the title, its presence in the abstract is checked and then in the keywords. When the study is not found in any field, the possibility of calibration of the search string is analyzed. 
In addition, during the validation process, we calculated the number of studies returned by each term to observe the string's evolution and terms saturation.

\subsection{Validation process results -- Step 1}
\label{validationresult1}

We executed the validation process defined in Section \ref{valprocess1}. Its results is presented in Table \ref{tbl:Calibration_Abstract}. For more detailed results see Table IV - in the Supplementary Tables document\footnotemark[\value{footnote}]. After analyzing the presence of string terms in the title of each study  we can conclude that our string was able to detect 1202 studies (78.2\%) out of the 1537 secondary studies from the control group (see Table \ref{tbl:Calibration_Abstract} column 2 -- line 26).

\begin{table*}
	\centering
	\footnotesize
	\caption{Presence of string terms in the title, abstract and keywords.}
	\label{tbl:Calibration_Abstract}
	\begin{tabular}{|p{2.7cm}|p{3cm}|p{3.4cm}|p{3.3cm}|p{1.5cm}|} \hline
		\textbf{ID} & \textbf{Studies retrieved by title} & \textbf{by title + abstract} & \textbf{by title + abstract + keywords} & \textbf{Not retrieved} \\ \hline
	
		Verner \textit{et al.}\@ & 20/24 (83.3\%) & 20/24 (83.3\%) & 20/24 (83.3\%) & 4 (16.7\%)\\ \hline
		Kitcheham and B. & 39/68 (57.3\%) & 59/68 (86.7\%) & 61/68 (89.7\%) & 7 (10.3\%) \\ \hline
		Bano \textit{et al.}\@ &  41/53 (77.3\%) & 52/53 (98.1\%) & 52/53 (98.1\%) & 1 (1.9\%) \\ \hline
		Zhou \textit{et al.}\@  & 93/110 (84.5\%) & 108/110 (98.2\%) & 109/110 (99.1\%) & 1 (0.9\%)\\ \hline 
		Goul\~ao \textit{et al.}\@ & 14/22 (63.64\%) & 19/22 (86.4\%) & 20/22 (90.9\%)  & 2 (9.1\%)\\ \hline 
		Nurdiani \textit{et al.}\@ & 26/41 (63.4\%) & 37/41 (90.2\%) & 38/41 (92.7\%) & 3 (7.3\%) \\ \hline 
		Napole\~ao \textit{et al.}\@ & 136/170 (78.9\%) & 167/170 (98.2\%)  & 170/170 (100\%) & 0 (0\%) \\ \hline 
		Hoda \textit{et al.}\@ &  22/28 (80.0\%) & 27/28 (96.4\%) & 27/28 (96.4\%) & 1 (3.6\%) \\ \hline 
		Marimuthu and C. & 55/60 (91.7\%) & 59/60 (98.3\%) & 59/60 (98.3\%) & 1 (1.7\%)\\ \hline 		
		Khan \textit{et al.}\@  & 19/24 (79.2\%) & 22/24 (91.7\%) & 23/24 (95.8\%) & 1 (4.2\%)\\ \hline 
		Budgen \textit{et al.}\@ & 34/37 (91.9\%) & 36/37 (97.3\%) & 36/37 (97.3\%) & 1 (2.7\%)\\ \hline 
		Singh \textit{et al.}\@ & 125/171 (73.1\%) & 167/171 (97.7\%) & 169/171 (98.8\%) & 4 (2.3\%)\\ \hline
		Rios \textit{et al.}\@ & 7/13 (53.8\%) & 12/13 (92.3\%) & 13/13 (100\%) & 0 (0\%) \\ \hline
		Ampatzoglou \textit{et al.}\@ & 153/165 (95.6\%) & 159/163 (96.4\%) & 159/163 (96.4\%) & 6 (3.6\%)\\ \hline
		Villalobos \textit{et al.}\@  & 8/22 (36.3\%) & 9/22 (40.9\%) & 9/22 (40.9\%) & 13 (59.1\%) \\ \hline
		Oliveira \textit{et al.}\@  & 2/4 (50.0\%) & 2/4 (50.0\%) & 2/4 (50.0\%) & 2 (50.0\%) \\ \hline
		Ruiz \textit{et al.}\@ & 1/5 (20.0\%) & 4/5 (80.0\%) & 4/5 (80.0\%) & 1 (20.0\%) \\ \hline
		Raatikainen \textit{et al.}\@ & 65/86 (75.6\%) & 81/86 (94.2\%) & 82/86 (95.3\%) & 4 (4.7\%)\\ \hline
		Delavari \textit{et al.}\@ & 55/94 (58.5\%) & 80/94 (85.1\%) & 84/94 (89.4\%) & 10 (10.6\%) \\ \hline
		Barros-justo \textit{et al.}\@  & 53/56 (94.6\%) & 54/56 (96.4\%) &  54/56 (96.4\%) & 2 (3.6\%)\\ \hline
		Curcio \textit{et al.}\@ & 11/14 (78.6\%) & 13/14 (92.9\%) & 13/14 (92.9\%) & 1 (7.1\%) \\ \hline
		Bedu \textit{et al.}\@ & 14/48 (29.17\%) & 18/48 (37.5\%) & 18/48 (37.5\%) & 30 (62.5\%)\\ \hline
		García-Mireles \textit{et al.}\@ & 11/12 (91.67\%) & 12/12 (100\%) & 12/12 (100\%) & 0 (0\%)\\ \hline
		Khan \textit{et al.}\@ (2) & 203/210 (96.7\%) & 203/210 (96.7\%) & 203/210 (96.7\%)  & 7 (3.3\%) \\ \hline
		
		\textbf{TOTAL (\textit{recall})} & \textbf{1202/1537 (78.2\%)} & \textbf{1422 (1202+220)/1537 (92.5\%)} & \textbf{1435 (1422+13)/1537 (93.4\%)} &  \textbf{102/1537 (6.6\%)} \\ \hline
	\end{tabular}

\end{table*} 

From the remaining 335 studies, we verified if a string' term were present in the abstract. A total of 220 additional studies were detected. 13 studies were retrieved by keywords, totaling 1435 of 1537 studies (1171+220+13 -- 93.4\%); 102 (6.6\%) studies were not retrieved by our proposed string. 

\begin{figure}[!h]
	\centering
	\includegraphics[width=0.8\linewidth]{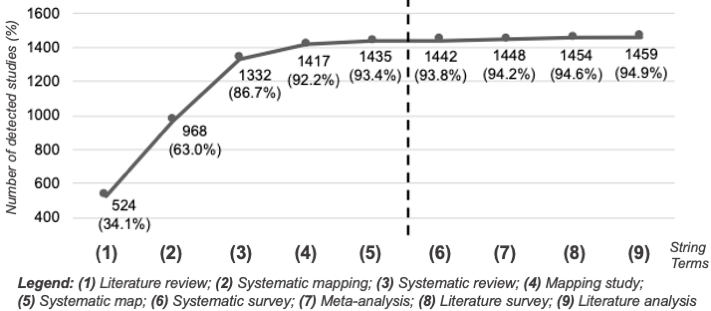}
	\caption{Number of studies detected by progressively adding terms to the search string. The dashed line represents the cutoff where additional terms provide an insufficient increase in \textit{recall}.}
	\label{string-evolution}
\end{figure}

Figure \ref{string-evolution} displays the evolution of the \textit{recall} value obtained by progressively adding more terms to the search string. For example, simply using ``literature review'' retrieves 34.1\% of the secondary studies on SE; adding ``systematic maping'' increases this fraction to 63.0\%, and so on. Our proposed search string (RQ1) is able to detect 93.4\% of the secondary studies on SE (see Figure \ref{string-evolution} left side of the dashed line).    

For didactic reasons, the validated search string terms (left side of the dashed line) in Figure \ref{string-evolution} were ordered according to the numbers of studies retrieved. Therefore, ``literature review'' was the term which returned the majority of the secondary studies (524 studies). All studies that contain the full name of the method ``systematic literature review'' was returned by ``literature review'' term as well. Next, ``systematic mapping'' is the second most term which detect secondary studies followed by ``systematic review" and ``mapping study'' terms.

We analyzed the remaining 102 studies that were not detected by our search string. Our goal in this step was to perform a search string calibration if it is reasonable.  We observed the title, abstract and keywords of each study, and identified how the authors described the adopted research method. We looked for terms that authors named the method adopted (secondary study). Our criteria in this step to consider a term, was that the named research method needs to repeat at least in two different secondary studies. As a result, we identified four new terms, they are: ``systematic survey'' (resulting in 7 more studies found), ``meta-analysis'' (6 studies), ``literature survey'' (6 studies) and ``literature analysis'' (5 studies).

We investigated in the literature to justify why secondary studies' authors called secondary studies these terms.  Our conclusions are described in the following.

\begin{table*} [bp]
	\centering
	\footnotesize
	\caption{General precision of the proposed search string}
	\label{tbl:val2}
	\begin{tabular}{|p{10cm}|p{1.1cm}|p{2cm}|p{1cm}|} \hline
		\textbf{String} & \textbf{Returned studies} & \textbf{Selected sec. studies in SE} & \textbf{Precision} \\ \hline  
		((``software engineering'') AND (``literature review'')) & 1294 & 643 & 49.7\%  \\ \hline 
		((``software engineering'') AND (``literature review'' OR ``systematic mapping'')) & 1695 & 985 (643+342) & 58.1\%  \\ \hline 
		((``software engineering'') AND (``literature review'' OR ``systematic mapping'' OR ``systematic review'')) & 1985 & 1144 (985+159) & 57.6\%  \\ \hline 
		((``software engineering'') AND (``literature review'' OR ``systematic mapping'' OR ``systematic review'' OR ``mapping study'')) & 2015 & 1166 (1144+22) & 57.8\%  \\ \hline
		((``software engineering'') AND (``literature review'' OR ``systematic mapping'' OR ``systematic review'' OR ``mapping study'' OR ``systematic map'')) & \textbf{2020} & \textbf{1169 (1166+3)} & \textbf{57.8\% } \\ \hline 
	
	\end{tabular}
\end{table*}

Regarding the terms ``systematic survey'' and ``literature survey'', one reason for using these terms in systematic reviews is the comprehensive definition of the term ``survey''. According to Pfleeger  \cite{Pfleeger95}, a survey is an empirical method that allows researchers to collect data from a large population aiming to generalize the findings. 
On the secondary studies case, the survey population is the included primary studies. However, in the case of tertiary studies, the single term ``survey'' by itself can add so many primary studies to the search result. In addition, calling a secondary study ``systematic survey'' or ``literature survey'' is a not recommended approach because it can make the detection and analysis of the studies retrieved by the search difficult. For example, the authors will need to investigate more deeply the paper to understand if the study is a secondary study in fact or not.

Regarding the term ``meta-analysis'', 
Cruzes \textit{et al.}\@ \cite{Cruzes11} conducted a tertiary study to assess the types and methods of research synthesis in secondary studies in SE. Only two reviews considered in their analysis, from the same research group, were classified as meta-analysis. In addition, our analysis shows that this scenario has changed over last years: the most recent study retrieved by the term ``meta-analysis'' from our control group was published in 2013. 

Finally the term ``literature analysis'' is defined as ``the practice of looking closely at small parts to see how they affect the whole'' \cite{Dawson05}. However, naming a study just as ``literature analysis" does not explicit if the study followed a systematic process \cite{Kitchenham15, Petersen15} during its conduction. This can lead to confusion or even misunderstanding during the decision of the inclusion or exclusion of a study.

As illustrated in Figure \ref{string-evolution}, adding the new terms to the search string during the calibration process, made string's total \textit{recall} value just increased in 1.5\% (from 93.4\% to 94.9\%) returning only new 24 secondary studies. As can be observed in Figure \ref{string-evolution}, after the cutoff the graph became almost linear, indicating saturation. Also, it is possible to observe that adding each term one by one the \textit{recall} increases less and less. For example,  the term ``literature survey'' to the existing string adds 0.4\% more studies (from 94.2\% to 94.6\%) and the term ``literature analysis''  adds a mere 0.3\% more studies (from 94.6\% to 94.9\%).

We examined the remaining 78 not-found secondary studies one by one. As a result, in some cases we found that the authors described as explanatory statements the adopted research method (type of study) in their abstract.  For example, ``Based on an analysis of the published literature\dots{}'', ``This research provides a preliminary review\dots{}'',``We provide an overview of the state-of-the-art\dots{}'', ``We carefully and systematically selected 98 articles\dots{}'', among others. In other cases, the authors mentioned the type of study explicitly (by name) or through explanatory statements only in the \textit{Introduction} or \textit{Methodology} sections of their paper. This fact reinforces the use of complementary searches added to automated searches; it also serves as a reminder to researchers that the research method should be explicitly mentioned in the title, abstract or keywords of the secondary studies they conduct.

Some specific results from the validation process execution can be highlighted. Before the calibration process execution, our proposed string was able to completely retrieve the set of secondary studies from three tertiary studies 
(Napole\~ao \textit{et al.}; Rios \textit{et al.}; and García-Mireles \textit{et al}.). 
However, if we add the new identified terms from our calibration process, this result is still the same. 

In summary, considering the terms analysis presented above as well as the low \textit{recall} increase, new terms have not been added to our proposed string during the calibration step. The validated suitable search string and search fields still the same mentioned as results from RQ1 and RQ2 respectively. 


The strings format must be verified for each DL. Such verification is necessary because the web search interfaces provided by these libraries continually change their rules, as also stated by Kuhrmann \textit{et al.} \cite{Kuhrmann17}. Table V in the Supplementary Tables document\footnotemark[\value{footnote}] presents different adaptations that were required in our search string to \textit{Scopus}, \textit{Web of Science} and \textit{IEEE Xplore} DLs considering search in the title, abstract and keywords of the studies.

\subsection{Validation process definition -- Step 2 }
\label{valprocess2}

We started the second step of our validation process aiming to use the same control group defined in Step 1, but it was not possible due to: (i) we do not have access to the list of returned studies by the DLs of each tertiary study; and (ii) we tried to replicate the search performed by tertiary studies from the control group considering exactly the reported search strings and time-frame, but we obtained widely discrepant results from the ones mentioned in the  tertiary studies. For this reason, we opted to calculate the \textit{precision} of the proposed string (after the first-step validation) running our proposed search string on the \textit{Scopus} DL, since it showed the most prominent DL in SE \cite{Mourao20}, and performing a selection procedure in order to identify which studies are indeed secondary studies in SE. The performed process is described following and its results are presented in Section \ref{validationresult2}. 

\textbf{-- Search string execution: } Our goal is to analyze the \textit{precision} of our search string in the SE context, for that we added the domain term "software engineering" considering just the first term of the first-step validation string ``literature review'' then run it at \textit{Scopus} DL and exporting the search results. We repeated this process adding progressively the string terms until the full proposed string (see Figure \ref{string-evolution}) has been considered. 

\textbf{-- Selection criteria:} Our selection procedure followed the following Inclusion (IC) and Exclusion (EC) criteria: IC1: The study must be in the SE context (follows the SWEBOK Guide areas \cite{swebok}); AND IC2: The study must mention explicitly the adoption of secondary study method (SLR or SM, or an update); OR IC3: The study must follow a known SE secondary studies guideline (such as\cite{Kitchenham15, Petersen15}); EC1: The study is not a secondary study e.g. it just uses results from a secondary study; OR EC2: Studies published before 2004 (before the first SLR guideline publication) OR EC3: The study is not published in English; OR EC4: The study was already considered (duplicated). 

\subsection{Validation process results -- Step 2 }
\label{validationresult2}

We followed the second-step validation process described in Section \ref{valprocess2}. The searches were performed in February 2021 and the selection procedure was performed by the first-author and revised by a second author (disagreements were solved by consensus). The final list of the selected studies is available online\footnotemark[\value{footnote}]. As can be seen in Table \ref{tbl:val2}, the results of our second-step validation process shows that the complete proposed string has a general \textbf{\textit{precision} of 57.8\%.} Also it is possible to notice that the highest obtained \textit{precision} was 58.1\% with the terms (``literature review'' OR ``systematic mapping''), but using only these two terms lead to miss 184 (over 15\%) secondary studies. Adding the term ``systematic map'', just more 3 secondary studies were found, not changing the \textit{precision} impact, but the quantity of studies to be analyzed increased just in 5 studies. In addition, as can be seen in Figure \ref{string-evolution} this term also has an impact on the completeness  results (\textit{recall}) of the string.

We also run a string containing the terms that resulted from our analysis before any validation (including terms after the dashed line in Figure \ref{string-evolution}). As result, 2457 studies were retrieved by \textit{Scopus} (an increase of 17.8\% in the number of studies to be analyzed), but no new secondary study was included. In this case, the \textit{precision} reduces to 54.1\%. In other words, adding more terms in the string only increase the reading loading effort spent in the selection process \cite{Kitchenham15}.

\section{Discussion} \label{discussions}
In this section, we discuss issues related to the results obtained in our analysis and we report on the main threats to validity of our study.

 An optimal search strategy is a search that strikes a balance between high \textit{recall} and high \textit{precision} \cite{Dieste07}. During the first-step validation process, four new possible terms were identified. However, after the two-steps analysis, we did not add these terms to our proposed search string.
As shows Figure \ref{string-evolution}, a string with five terms returns 93.4\% of the included studies and adding four more terms increase \textit{recall} only in 1.5\%. In addition, the \textit{precision} reduced after the addition of these four terms. 
In other words, \textbf{our results showed that adding new terms and building long search strings does not guarantee an efficient return of relevant studies nor a reduction of the effort during the manual filtering in the selection procedure.} 
A search string with more terms (in our case, connected by logical operator \textit{OR}) potentially leads to return a large number of candidate primary studies including many false positives \cite{Kitchenham15}. In addition, is worth mentioning that some DLs impose limitations on the number of characters or terms, for example, \textit{IEEEXplore} limits in 15 the number of terms in the search \cite{Singh18preprint}. These limitations lead researchers to have to divide the search string into smaller parts, demanding multiple searches and deletion of repeated studies. In fact, an efficient search string with few terms is the goal. Simple queries are usually better accepted by DLs \cite{Kuhrmann17}.

According to the search strategies scales used for evaluating search terms proposed in \cite{Dieste07}, a optimum search strategy has a \textit{recall} rate between 80-99\% and a max \textit{precision} rate between 25-60\%. Our results demonstrate the most significant possible balance when compared with these metrics: 93.4\% of \textit{recall} with 57.8\% of \textit{precision}. 

\begin{figure}[!h]
	\centering
	\includegraphics[width=0.8\linewidth]{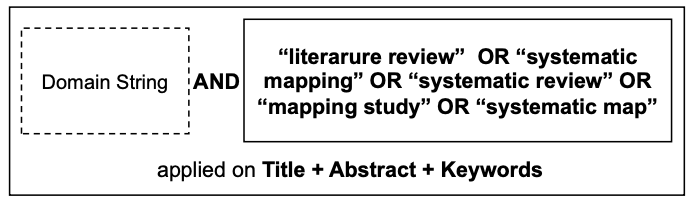}
	\caption{A suitable search string and preferable fields for searching secondary studies (final version).}
	\label{conclusion}
\end{figure}

As illustrated in Figure \ref{conclusion}, a search string for searching secondary studies should be formed by: (1) terms and their synonyms related to research domain connected by \textit{OR} logical operator; (2) logical operator \textit{AND}; (3) terms and their synonyms related to secondary studies connected by \textit{OR} logical operator. In summary, the synonyms suggested by our study to detect secondary studies are:  \textbf{``literature review'' \textit{OR} ``systematic mapping'' \textit{OR} ``systematic review'' \textit{OR} ``mapping study'' \textit{OR} ``systematic map''.}  After the search string elaboration, it must be applied in \textbf{title}, \textbf{abstract} and \textbf{keywords}. 

Elaboration of a standard string for the SE areas and subareas is not a trivial task and it is not possible to guarantee that the synonyms contained in the selected domain will return all the necessary evidence to answer the research questions proposed by a given tertiary study. Due to this fact, the definition of terms for search strings' domain was not addressed in this study.

One point to be discussed is the search execution in title, abstract and keywords fields. Our study shows that apply the proposed search string in these fields leads to a high \textit{recall} of the studies. However, not always the adopted search method (secondary studies) is described in these fields requiring a search in the full text of research papers, which is not supported by indexing services nor all publishers' DLs \cite{Kitchenham15}. 


Our results demonstrate the most adopted terms related to SLRs and SMs (higher \textit{recall}). Therefore, we suggest that the titles of SE secondary studies be formed of two parts, following the template ``research topic: research methodology''. 
For example: ``Knowledge Management Practices in GSD: A Systematic Literature Review'' \cite{Arshad12}.

Our research also showed that abstracts were the field that most returned secondary studies (See Table \ref{tbl:presence} -- line 14, column 3). Therefore, researchers must be careful at explaining the research method in the abstract. In addition, it is important to verify if the terms that are being used for describing the method or methodology addressed are straightforward. For example, writing ``We performed a systematic analysis of the existing literature following established guidelines in SE'', rather than simply ``We performed a systematic review''.

The string proposed in this study identified 93.4\% of the secondary studies in SE with a \textit{precision} of 57.8\%. According to Petersen \textit{et al.}\@ \cite{Petersen15} and Kitchenham \textit{et al.} \cite{Kitchenham15}, a researcher can obtain more accurate results by combining an automated search strategy with complementary searches, such as \textit{Snowballing} \cite{Wohlin14} or manual search. Considering that both complementary search strategies are independent of the search terms, they may be used as complementary searches during the conduction of a tertiary study.

\subsection{Threats to Validity} \label{threats}
We report on the main threats to validity of our study and the adopted mitigation strategies: (1) \textbf{Construct validity.} The main limitation of this study is that we had no access to the list of secondary studies returned by the automated search performed by authors. For this reason, we made use of the lists of the included secondary studies that are available. We assumed that the other studies were excluded because they were not relevant. We also acknowledge that the comparison of our \textit{recall} results was made using a control group that is human-derived and it most likely did not achieve 100\% \textit{recall} either. Thus, as a first-cut assessment, we believe our study met its goal. The difficulty in elaborating search strings is also related to particularities of the domain under consideration and the expertise needed to judge upon the relevancy \cite{Kuhrmann17}. Consequently, we opted to limit our analysis only in the secondary studies' related terms and for the validation process -- step 2, we considered as SE studies the studies related to the areas mentioned in the SWEBOK guide \cite{swebok}. 
(2) \textbf{External validity.} It is important to observe the limitation existing in relation to the sample of studies analyzed. The terms of our string were initially extracted from seven tertiary studies \cite{Garousi16}. However, we constructed a control group of 1537 secondary studies based on 24 other tertiary studies. In total, our analysis considered 31 (7+24) different tertiary studies. In addition, we executed the proposed search string in a well-known DL and analyzed its results systematically. The study design and the validation process were rigorously performed.
\section{Conclusions}\label{conclusions}

This study aimed to assist investigations in SE area to create a search string to find secondary studies. In light of our findings, the main contributions of this research are: (1) A suitable search string that can efficiently retrieve secondary studies wherein five search terms were able to reach over 90\% of  \textit{recall} and a general  \textit{precision}  of  almost  60\%; (2) The proposed suitable search string may help SE researchers detect secondary studies, possibly enabling time-saving in the search and selection procedures, especially for novices researchers that usually are not familiar with secondary studies terminologies; and (3) We demonstrated that the preferred fields to search for secondary studies are title, abstract and keywords. 
	
As future work, we intend to further investigate search strategies as a whole, including a deeper investigation on automated search strategy, mainly on its combination with other search methods and on the selection of DLs. In addition, we intend to develop a tool to (semi) automatically adapt the search strings in DLs for SE studies.


\bibliographystyle{unsrt}  
\bibliography{references}  

\end{document}